\documentclass{iopconfser}

\usepackage{siunitx}
\usepackage{color}
\usepackage{mathtools}
\usepackage[normalem]{ulem}
\usepackage{adjustbox}
\usepackage{array}
\usepackage{booktabs}
\usepackage{multirow}
\usepackage{amsmath}
\usepackage{url}
\usepackage{natbib}
\usepackage{float}
\usepackage{natbib}

\begin{document}

\title{WOFRY: a package for partially coherent beamline simulations in fourth-generation storage rings}

\author{Manuel Sanchez del Rio$^{1}$, Juan Reyes-Herrera$^{1}$, Rafael Celestre$^{2}$ and Luca Rebuffi$^{3}$}

\affil{$^1$European Synchrotron Radiation Facility, 71 Avenue des Martyrs, 38000 Grenoble, France}

\affil{$^2$Synchrotron SOLEIL, L'Orme des Merisiers Départementale 128, 91190 Saint-Aubin, France}

\affil{$^3$Argonne National Laboratory, 9700 South Cass Avenue, Lemont, Illinois 60439, USA}

\email{srio@esrf.eu}

\begin{abstract}
We present WOFRY (Wave Optics FRamework in pYthon), a specialized toolbox designed for wave optics modeling, with particular emphasis on partial coherence. This package is tailored to assist synchrotron scientists and engineers in the prototyping and modeling of X-ray sources and optics. WOFRY offers functionalities for generating and propagating 1D and 2D wavefronts, along with various tools for interacting with different optical elements.
The software developed is available in the OASYS \citep{OASYS} suite.
%
\end{abstract}


\section{Introduction}

Wavefront simulations are getting more and more interest in 4th generation synchrotron facilities for simulating optical devices and instruments. Fully coherent sources, or sources with very high coherence fraction can be represented by a single wavefront. However, most 4th generation storage-ring-based sources working in the X-ray regime are not completely coherent, with coherence fraction of the order of 10\% or less. Therefore partial coherence models are necessary, representing the source with multiple wavefronts. 

Several codes exist in the synchrotron community for physical optics calculations, like SRW \citep{codeSRW}, XRT \citep{XRT}, WISE \citep{wise}, MOI \citep{MOI}, etc. with different implementations, each featuring pros and cons, but none of them satisfy all our requirements. The objectives that drove us to develop WOFRY are i) the need of a fast a quick prototyping of a complete beamline using physical optics, ii) run it in a powerful and simple user interface, iii) light enough to run in a laptop, iv) model accurately the partial coherence (including coherent fraction along the beamline), v) exact simulation the surface errors in mirrors and lenses, and vi) automatic generation of scripts that could be used in cloud computing.    

WOFRY is built upon a framework library \citep{syned} featuring key components such as the ``Propagation Manager" and ``Propagator" objects, which serve as the interface to wavefront propagation algorithms, providing a high-level representation of the propagation mechanism of wavefronts.
In terms of source simulation, WOFRY goes beyond basic plane, spherical and Gaussian waves. It makes calculation of partial coherence using not only the Gaussian Shell-model but also performs coherent mode decomposition of undulator radiation \citep{SanchezdelRio2022CMD}.
Several optical elements are implemented, including absorbers (slits and stops), reflectors (mirrors and gratings), and refractors (ideal lenses, real lenses, compound refractive lenses, and thin objects \citep{Celestre2019}). The physical models and algorithms implemented in the different components of WOFRY are presented in this paper.


\section{WOFRY wavefronts and their propagation in free space}


We use wavefronts that are defined in a plane perpendicular to the main propagation direction. We define the propagation direction $y$. In a 2D model, the plane has coordinates $(x,z)$ (horizontal and vertical). In a 1D mode, there is only one spatial coordinate $x$ that can be either horizontal or vertical. 
The wavefront is represented by the electric field $E$ of a monochromatic component (wavelength $\lambda$, wavenumber $k = 2 \pi / \lambda$ or photon energy $\omega$) at a given position $y$ along the propagation direction (coincident with the optical axis).
This electric field or electric disturbance is a complex scalar $E(x;y=0,\omega)$. It can be also expressed by two real numbers, and amplitude $A$ and a phase $\phi$,  $E= A \exp{(i\phi)}$. In the case of polarized beams, two complex scalars, $E_\sigma$ and $E_\pi$ are used for $\sigma$ and $\pi$ polarization, respectively.
The wavefront intensity is the square of the modulus of the amplitude: $I=|E|^2=|E_\sigma|^2 + |E_\pi|^2$. In WOFRY, the wavefront is an array of complex numbers over a discretized mesh. 1D wavefronts are $E(x_i;y,\omega)$ and 2D wavefronts are $E(x_i,z_j;y,\omega)$. Every wavefront is defined at a position $y$ and for a photon energy $\omega$. This dependency is not explicit in the following.

The wavefront evolves when transported in free space. A propagator computes a wavefront $E(x;y_2)$ in a point $y_2$ knowing the wavefront in another position $E(x;y_1)$ and the distance $\Delta$ in free space $y_2 = y_1 + \Delta$. WOFRY has implemented two propagators, described here.

\subsection{Direct implementation of integral Rayleigh-Sommerfeld propagator}
\label{sec:integralPropagator}

The Rayleigh-Sommerfeld propagator for small-angle approximation expresses the electric field at a spatial point $\vec{r}'$ as an integral of the electric field at a spatial point $\vec{r}$ \citep{goodmanfourier}
\begin{equation}\label{eq:RSpropagator}
E(\vec{r}') =  \frac{k}{2 \pi i} \int \frac{E(\vec{r})}{|\vec{r}'-\vec{r}|} e^{ i k |\vec{r}' - \vec{r}|  }  d\Sigma,
\end{equation}
where the integral is made over the domain of the source (the surface $\Sigma$). 
This propagator can be applied to numeric discrete 1D wavefronts, and the integral reduces to a sum
\begin{equation}\label{eq:discreteRSpropagator}
E(x_2;y_2) = \frac{e^{i k \Delta}}{\sqrt{2 \pi i \Delta}}  \sum_{i=0}^{N-1}  E(x_{1,i};y_1) e^{i k \sqrt{(x_2 - x_{1,i})^2 + \Delta^2} }.
\end{equation}
Note that the sum extends over the $N$ points of the sampled incident wavefront has to be done for each coordinate at the transported wavefront, thus the number of operations is of the order $N^2$. This simple propagator gives flexibility to define different gridding and limits permitting the adjustment of the spatial domain or window and spatial resolution when working with converging or divergent wavefronts. Moreover, the propagated wavefront can be computed in a plane (or line) non parallel to the incident one, a feature that is exploited in section~\ref{sec:mirror1D}.

\subsection{Fresnel propagator using FFT: The zoom propagator}
\label{sec:zoomPropagator}

The Fresnel propagator is obtained by making a Taylor expansion of the quadratic phase in Eq.~\ref{eq:RSpropagator}. In 1D is 
\begin{equation}\label{eq:fresnelPropagator}
E(x;y_2) =  \frac{e^{i k \Delta}}{\sqrt{i \lambda \Delta}} \int E(x';y_1) e^{ \frac{i k}{2 \Delta}  (x-x')^2  }  dx'.
\end{equation}
The Fresnel integral propagator can be seen as convolution of the wavefield with a Gaussian kernel. One can write Eq. (\ref{eq:fresnelPropagator}) in convolution form \citep{goodmanfourier}, involving a Fourier transform $\mathcal{F}$ and an inverse Fourier transform $\mathcal{F}^{-1}$. The real benefit of using this scheme comes from the use of Fast Fourier Transforms, that reduce the number of operations from $N^2$ to $N \log_2 N$. Its use is essential when doing simulations in 2D, because the direct calculation of the integral lead to $N^4$ operations at the limit of calculation power of usual laptop computers.
It is possible to calculate the propagated field in a ``zoomed'' window, thus permitting optimizing the wavefront sampling in cases for propagating highly convergent or divergent wavefronts. This is implemented using the method proposed in \citep{schmidt}. The problem reduces to a convolution problem of the unpropagated field field $E(x;y_0)$ affected by a phase $P$ with a kernel $K$, and the result affected by a global phase $P^G$
%
%
The zoom propagator in 1D can be expressed 
\begin{equation}
E(x; y_2) = P^G \mathcal{F}^{-1} \Big\{ \mathcal{F} \big\{ E(x;y_1)~~P \big\} K \Big\},
\end{equation}
where
\begin{equation}
P^G =  \frac { e^{ik\Delta }}{\sqrt{m_x} }e^{i \frac{k}{2 \Delta} \frac{m_x - 1}{m_x}x_2^2},
\end{equation}
\begin{equation}
P = e^{i \frac{k}{2\Delta} (1-m_x)x_1^2 },
\end{equation}
\begin{equation}
K = e^{-i \pi \lambda \Delta \frac{u^2}{m_x} }.
\end{equation}
The term $m_x$ is the magnification (zoom) factor. WOFRY also implements the zoom propagator in 2D, with different magnifications in horizontal and vertical directions, described in \citep{pirro}.

\section{Models for the Source widgets}

\begin{figure}
    \centering
    \includegraphics[width=0.29\textwidth]{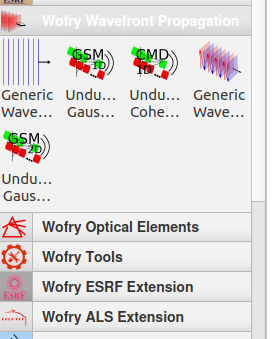}
    \includegraphics[width=0.29\textwidth]
    {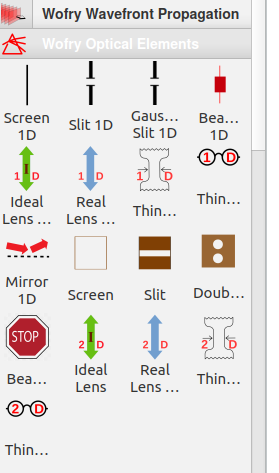}
        
    \caption{The WOFRY menus for the sources (left) and optical elements (right).}
    \label{fig:oasysmenus}
\end{figure}

We describe here the models for the different widgets that are used to create sources, available in the menu shown in Fig.~\ref{fig:oasysmenus}.

\subsection{``Generic Wavefront" widgets}

In this section, we present the models for the different sources available in the ``Generic Wavefront 1D" and ``Generic Wavefront 2D" widgets. 
We restrict the formulation here to a 1D model. For the 2D models, the wavefront is created typically by composing the horizontal and vertical 1D models.  

\subsubsection{Plane Wave}\hspace*{\fill} \\
The simplest wavefront is a plane wave, with constant complex amplitude for any $x$ coordinate: 
\begin{equation}
   E(x;y=0,\omega)=E_0=A_0 e^{i \phi_0},
\end{equation}
where $E_0$ is a complex value that can be expressed in its constant real amplitude $A_0$ and a constant phase $\phi_0$. A plane wave is infinite along the $x$ direction. However, when representing a wavefront spatially, in simulation, the electric field has to be sampled over an array of discrete values of complex amplitude, and they are necessarily defined over a finite $x$ interval. 

\subsubsection{Spherical Wave}\hspace*{\fill} \\
A spherical wave (strictly speaking a circular wave in 1D, but we keep the terminology used in 2D) emanates from a fixed point and has a constant complex amplitude over a sphere of a given radius $R$. Obviously it cannot be represented at the source point ($y=0$, center of the sphere) and our wavefront must be sampled at a given distance $y=R$ and over a line perpendicular to the radius and tangent to the sphere. At $x=0$ (a point in the sphere) the field has a constant value equal to the value at the surface. But at a $x{\ne}0$, over a range of $x$-values $x{\ll}R$ (i.e. for small numerical apertures, NA) the distance from $x$ to the sphere parallel to the $y$ direction gives an optical path that modifies the phase by $k ~\Delta x$. It is easy to deduce that the wavefront in the line tangent to the sphere has the expression
\begin{equation}
\label{eq:sphericalWave}
    E(x;y=R,\omega)  = E_0 e^{i k x^2 / (2 R)}.
\end{equation}

\subsubsection{Gaussian wave}\hspace*{\fill} \\
A Gaussian beam at the source position $y$=0 has constant phase and intensity following a Gaussian distribution with standard deviation $\sigma_I$, thus the electric disturbance is: 
\begin{equation}
\label{eq:gaussianSource}
    E(x;y=0,\omega) = E_0 e^{-x^2 / (4 \sigma_I^2)}
\end{equation}

\subsubsection{The Gaussian Shell-model}\hspace*{\fill} \\
In the Gaussian Shell-model (GSM) the cross-spectral density (CSD) is, 
\begin{equation}
W(x_1,x_2;\omega) = A^2 e^{-(x_1^2+x_2^2)/(4 \sigma_I^2)} e^{-(x_2-x_1)^2/(2 \sigma_{\mu}^2)},
\label{GS_CSD}
\end{equation} 
with $A$ a constant, and $\sigma_I$ and $\sigma_\mu$ are related to ``intensity" and  ``coherence" widths. All are positive constants. 

In Optics, the expansion of a generic CSD in a series of orthonormal polynomials is called coherent mode decomposition (CMD)  
\citep{mandel_wolf}
\begin{equation}
W(x_1,x_2;\omega) = \sum_{n=0}^{\infty} \lambda_n(\omega) \Phi_n^*(x_1;\omega) \Phi_n(x_2;\omega). 
\label{CMD}
\end{equation}
where $\lambda_n$ are the eigenvalues and $\Phi$ the eigenfunctions (coherent modes). 
Note that Eq.~(\ref{CMD}) can be written as 
\begin{equation}
W(x_1,x_2,\omega) = \sum_{n=0}^{\infty} E_n^*(x_1,\omega) E_n(x_2,\omega), 
\end{equation} with $E_n(x, \omega) = \sqrt{\lambda_n(\omega)} \Phi_n(x,\omega)$, and intensity (spectral density) of the i$^\text{th}$-wavefront $W_i(x, x, \omega) =  |E_i(x, \omega)|^2$.
The eigenvalues $\lambda_n$ are a measure of the intensity.
The occupancy $\eta$ of the i-th mode is defined as the ratio of its intensity to the total intensity 
\begin{equation}
\eta_i(\omega) = \frac{\lambda_i(\omega)}{\sum_{n=0}^{\infty} \lambda_n(\omega)}.
\end{equation}
The Coherent Fraction ($CF$) is the occupancy of the first coherent mode $CF=\eta_0$.

The CSD of the GSM in Eq.~(\ref{GS_CSD}) can be expanded analytically in coherent modes \citep{Starikov82}. The resulting eigenvalues are
\begin{align}
\lambda_0 &= A \sqrt{\pi/( a+b+c)}; \\ 
\lambda_n &= \lambda_0 q ^n,
\end{align}
with $a = (4 \sigma_I^2)^{-1}$, $ 
b = (2 \sigma_{\mu}^2)^{-1}$, $ 
c = (a^2 + 2 a b)^{1/2}$,
$q = [1 + \beta^2/2 + \beta\sqrt{(\beta/2)^2+1}]^{-1}$ 
and $\beta=\sigma_{\mu}/\sigma_I$.
The eigenfunctions are
\begin{equation}
\Phi_n(x) = \left( \frac{2c}{\pi} \right)^{1/4} \frac{1}{\sqrt{2^n n!}} H_n(x\sqrt{2c})e^{-cx^2}
\label{GSeigenvalues}
\end{equation}
with $H_n$ are the physicist's Hermite polynomials of order $n$. 

\begin{figure}
    \centering
    \includegraphics[width=0.69\textwidth]{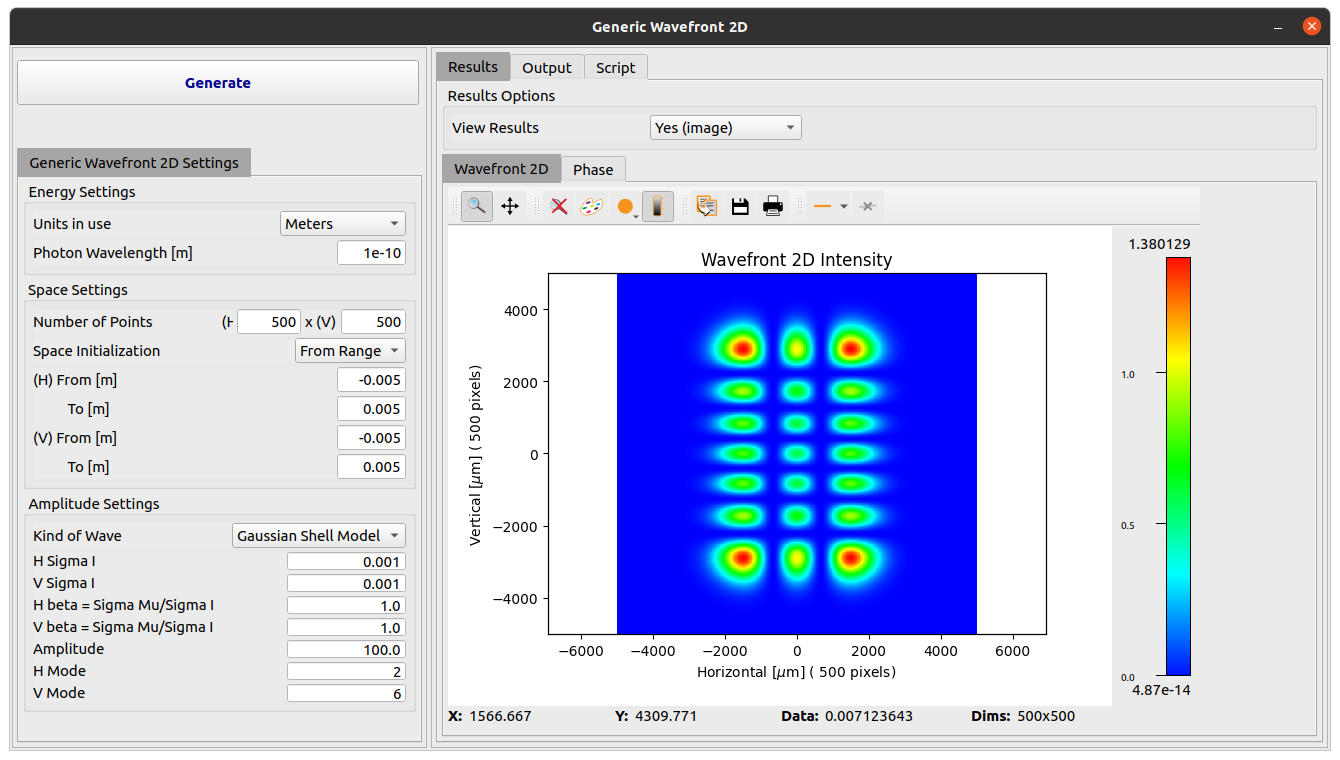}
        
    \caption{Example of a 2D GSM source created with the ``Generic Wavefront 2D" widget.}
    \label{fig:GSM2D}
\end{figure}

Note that the 0$^\text{th}$ mode is $E_0(x) \propto \exp(-c x^2) = \exp(-\sqrt{(16 \sigma_I^2)^{-1} + (4 \sigma_I^2 \beta^2)^{-1}} x^2)$ which in general differs from the basic Gaussian wavefront in Eq.~(\ref{eq:gaussianSource}), but both approximate well when $\beta \gg 1$. An example of GSM 2D source combining 2$^\text{th}$ order mode in horizontal and 6$^\text{th}$ order mode in vertical is shown in Fig.~\ref{fig:GSM2D}.
The occupancy of coherent modes for the GSM is   
\begin{equation}\label{eq:GSMoccupancy}
\eta_i = q^i(1-q), 
\end{equation}
thus the coherent fraction is 
\begin{equation}\label{eq:GSMcoherentfraction}
    CF=\eta_0=1-q,
\end{equation}
and the cumulated occupancy
\begin{equation}\label{eq:GSMcumulatedoccupancy}
\ \sum_{n=0}^{i-1} \eta_n = 1-q^i.
\end{equation}
From here one can calculate the number of modes needed to reach a cumulated occupancy $c_0$, which is $n=\ln(1-c_0)/\ln q$.

\subsection{``Undulator Coherent Mode Decomposition 1D" widget}
\label{sec:CMD}

A full model for coherent mode decomposition in 2D was proposed and implemented in the COMSYL code \citep{glass2017}. The calculation of the CSD and its expansion in normal modes is a complex and heavy task that requires long calculations in supercomputers. 
An important reduction in the calculation effort is obtained with 1D model of the waves. In this model we suppose that the CSD is separable in its 1D horizontal $x$ and vertical $z$ coordinates, therefore the $W_{2D}$ becomes a product of two CSD functions $W$ of two variables for a given photon frequency $\omega$:
\begin{equation}
W_{2D}(x_1,z_1,x_2,z_2;\omega) = W(x_1,x_2;\omega) W(z_1,z_2;\omega).
\label{eq:CSD_2D}
\end{equation}
Each $W$ function is treated separately in a similar way affecting the $x$ (horizontal) and $z$ (vertical) coordinates.
The eigenvalues and the coherent modes result from the diagonalization of the CSD.  In 1D, the CSD is a function of two variables, represented in a matrix that can be easily diagonalized using standard tools availables in python-numpy. 
The 1D method  available in the WOFRY widget ``Undulator Coherent Mode Decomposition 1D" has been described and extensively tested in \citep{SanchezdelRio2022CMD}.

\subsection{``Undulator Gaussian 1D" widget}
\label{sec:undulatorG}

This aims to create a wavefront that ``approximates" the coherent emission (zero emittance) of an undulator at resonance.
A single electron (or filament beam) traveling in an undulator with a pure sinusoidal magnetic field in one direction will produce a complicated wavefront with geometry that varies as a function of the photon frequency (see, e.g., \citep{elleaume}). At the resonance energy the angular emission in the far field can be approximated by a Gaussian function of width \citep{elleaume}
\begin{equation}
\label{eq:undulatorDivergence}
    \sigma'_u \approx\sqrt{\frac{\lambda}{2 L}},
\end{equation}
with wavelength of the photons at the undulator resonance, $\lambda$, and undulator length, $L$. At the source point, the size of the source can also be approximated by a Gaussian of width
\begin{equation}
\label{eq:undulatorSize}
    \sigma_u \approx\sqrt{\frac{\lambda L}{2 \pi^2}} = \frac{\lambda}{2 \pi} \frac{1}{\sigma'}.
\end{equation}

An important remark is that for undulator radiation $\sigma_u \sigma'_{u} \approx (2\pi)^{-1}$ whereas for a Gaussian beam, the propagation of Eq.~(\ref{eq:gaussianSource}) in the far field has a width in intensity $\sigma'_{I}$ that verifies $\sigma_I \sigma'_{I} \approx (4\pi)^{-1}$. This is just one reason why undulator beams differ from Gaussian beams. If we match the undulator beam with a Gaussian beam at the source position, they will differ in the far field. Contrarily, if we approximate the undulator far-field beam with a Gaussian beam, the propagation of these two wavefronts at the source position will differ. 
Consequently, we must make a choice, and approximate the undulator at either i) the source position, or ii) the screen position. 
For the first case we consider a  Gaussian source as defined in Eq.~(\ref{eq:gaussianSource}) with $\sigma_I=\sigma_u$.
In the second case, the far field can be approximated by a spherical wave Eq.~(\ref{eq:sphericalWave}) with origin in the undulator center modulated with an amplitude that follows the Gaussian in Eq.~(\ref{eq:undulatorDivergence}), therefore
\begin{equation}  \label{eq:undulatorBySphericalWave}
    E(x;y=y_0,\lambda) = E_0 e^{i k x^2 / (2 y_0)} e^{-x^2/(4 \sigma'_u{}^2 y_0^2)}.
\end{equation}

For practical purposes, the Gaussian source approximation can be used for optical systems accepting a high NA, whereas for systems of small NA, the spherical wave approximation may be preferred. A widget ``Undulator Gaussian 1D" is available.

\subsection{``Undulator Gaussian Shell-model 1D" widget}
\label{sec:undulatorGSM}

The real emission of the undulator is is very different than a Gaussian. However, the (inadequate) use of a Gaussian approximation for the undulator radiation is often found in literature, e.g.  \citep{coisson1997}.
We have developed the ``Undulator Gaussian Shell-model 1D" widget
aiming to compare the CMD of the undulator emission discussed before with the approximation given by GSM.

For representing the undulator radiation by a GSM, we need to obtain the $\sigma_I$ and $\sigma_\mu$ (or $\beta$) values. The intensity $\sigma_I$ matches the intensity distribution of the undulator source that is approximated as the convolution of the electron bunch size $\sigma_x$ with the radiation cross section $\sigma_u$. The other parameter $\beta$ is obtained by matching the GSM coherence fraction in Eq.~(\ref{eq:GSMcoherentfraction}) with the undulator coherence fraction approximated by
\begin{equation}\label{eq:CFphasespace}
    CF_u = \frac{\sigma_u\sigma_{u'}}{\sqrt{\sigma_x^2+\sigma_u^2}\sqrt{\sigma_{x'}^2+\sigma_{u'}^2}}.
\end{equation}
The resulting value for GSM $\beta$ that approximates the undulator emission is $\beta=CF_u/\sqrt{1-CF_u}$ (obtained from Eq.~\ref{eq:GSMcoherentfraction}).

There many arguments to prefer modeling undulators by using numeric CMD (section \ref{sec:CMD}) instead of the approximation by GSM. For instance, the small differences in a mode in terms of shape (even maintaining the same FWHM) may give to completely different results after propagation. Moreover, the GSM modes are real functions (constant phase) whereas the modes from CMD of undulator sources are complex functions (with a non-constant phase). A comparison between the 1D undulator CMD and this approximation of  undulator radiation by GSM is in section~\ref{sec:CMDvsGSM}


\section{Models for the beamline element widgets}
\label{sec:elements}

In cases where the distance between elements is significantly greater than the mirror sizes, the optical elements can be approximated as thin elements (zero thickness along the $y$ axis), and their effect can be encapsulated in a complex transmission amplitude,
\begin{equation}
    \label{eq:thinelement}
    R(x;\omega)=r(x;\omega) e^{i \rho(x,\omega)}.
\end{equation}
Therefore, the wavefield after an optical element placed at position $y=y_0$ will be the wavefield at $y_0$, just before the interaction, multiplied by the complex transmission:
\begin{equation}
    E'(x;y=y_0,\omega) = E(x;y=y_0,\omega) R(x;\omega)
\end{equation}

\subsection{Slits and beam stoppers}

A generic aperture is a mask that transmits a part of the wavefront in a range $[x_{min},x_{max}]$ and absorbs the rest. It can be
\begin{equation}
R(x;\omega) =
\left\{
\begin{matrix}
A,  & \mbox{~~if~~} x_{min} \le x \le x_{max};
\\ 
1 - A, & \mbox{~~elsewhere}.
\end{matrix}
\right.
\end{equation}
When the element is a slit, then $A=1$. If it is a beamstop, then $A=0$. For an aperture of finite length $a$ centered with the beam, $x_{min}=-a/2$ and $x_{max}=a/2$.

A useful widget is the ``Gaussian Slit 1D" where $A=\exp(-x^2/(4\sigma_a))$ (note the factor 4 is because $\sigma_a$ is related to the intensity). The aperture acts as a Gaussian appodization window. Although artificial, it is interesting to shape the beam with a Gaussian profile, which remains a Gaussian after propagation. Therefore, a Gaussian slit acting on a plane wavefront does not produce diffraction fringes. 
The slit $a$ aperture matches the FWHM, i.e. $\sigma_a=a/2.355$.

\subsection{Ideal lenses}
The ``Ideal Lens 1D" and ``Ideal Lens 2D" are useful widgets for defining a first prototype of a beamline because they can model approximately the effect of any focusing device, like mirror or lens. Later, in a refined simulation, they are replaced by other widgets with more accurate models. 
An ideal lens converts a plane wave into a spherical wave collapsing to the focus at a distance $f$ downstream from its position.
Therefore $R(x;\omega) = \exp[-i k~x^2/(2 f)]$.
An ideal coupling of $N$ ideal lenses will present a focal length $f=(f_1^{-1}+f_2^{-1}+...+f_N^{-1})^{-1}$. 

\subsection{Thin objects}\label{sec:thinobject} 
The ``Thin Object 1D" and ``Thin Object 2D" widgets permit to compute the modification of a wavefron (thus the absorption and focusing effects) by a layer of material (placed perpendicular to the beam) that is shaped with a height profile $h(x)$ in 1D ($h(x,z)$ in 2D). The height profile is given in a mesh file. This is useful to add surface errors to lenses or other transmission elements, but they can also be used to define a full lens, a corrector, or an element with arbitrary profile.    

A refractor made by a material with refraction index $n(\omega)=1-\delta(\omega)+i\beta(\omega)$ 
and thickness profile $h(x)$ adds a phase to the wavefront $-\lambda \delta(\omega) h(x)$ and reduce its amplitude by $\exp(-\mu h(x)/2)$, where $\mu=(4 \pi/\lambda) \delta(\omega)$ is the (intensity) attenuation coefficient and $\lambda$ the photon wavelength. These expressions can be applied to profile $h(z)$ assuming valid the thin object approximation, i.e., assuming that the object is thin enough to avoid multiple scattering and diffraction in its bulk. 

\subsection{Real lenses}
The changes induced by a real lens in the wavefront can be calculated using the thin object approximation discussed below, using the lens profile. Usually, a single lens has a parabolic profile $h(x)=x^2/(4R)$ with $R$ the radius of the apex. A lens has two parabolic profiles (front and back) separated by a lens thickness $d$ and a flat profile outside the lens aperture $A$, therefore:
\begin{align}
    h(x) &= \frac{1}{2R} x^2 + d; & x < A/2\\ \nonumber
    h(x) &= \frac{1}{2R} (A/2)^2 + d; & x \ge A/2.
\end{align}
The widget ``Real Lens 1D" uses this profile and also permitting to add a constant thickness,  replicate the profile (to model $N$ lenses in a CRL), and add a surface error profile. A similar bi-dimensional model is used in the ``Real Lens 2D" widget.  

\subsection{Refractive correctors}
\label{sec:refractorCorrector}

A corrector is an optical element designed to suppress specific optical aberrations in the wavefront. The purpose of the  ``Thin Object Corrector 1D" widget (a similar concept for 2D)  is to calculate the profile $h(x)$ of a thin object that allows to focus an arbitrary incoming wavefront to a given point (waist) at a distance $D$ downstream from the corrector. In other words, this object would transform the incoming wavefront into a spherical (circular in 1D) wavefront collapsing to the focus. The profile of such refractive ``corrector"  can be calculated using the phase difference $\Delta\phi(x)=\phi_S(x)-\phi(x)$ where $\phi_S=k x^2 / (2 D_w)$ is the phase of the circular wave collapsing at $D_w$, $k$ is the wavenumber ($k=2\pi/\lambda$) and $\phi$ is the phase of the incoming wavefront. The corrective profile is $z(x)=-\Delta\phi/(k \delta(\omega))$.

\subsection{``Mirror 1D" widget}
\label{sec:mirror1D}

This widget permits to compute the effect of a grazing incidence mirror with different shapes (flat, circle, ellipse, parabola) also affected by a surface error profile.
Let us consider a perfectly reflecting surface (no absorption) with a profile $h(w)$ with $h$ the elevation (height) and $w$ the linear coordinate in a reference frame attached to the reflector with origin in the reflector's center. A plane reflector has $h(w)=0$ and, for example, a circular mirror of radius $R$ has $h(w)=R-\sqrt{R^2 - w^2}$. The profile $h(w)$ can also match a deformation originated for instance by heat load or a mirror surface error.
Note that for a grazing incidence mirror  the  {\it thin object} approximation is not applicable. The solution implemented in WOFRY consists in using the integral propagator (Sec.~\ref{sec:integralPropagator}).
Let us represent the mirror surface coordinates as $(w,h)$. Let be $p$ the distance from the wavefront $E_0(x,;y=0,\omega)$ to the center of the mirror placed at a grazing incidence $\theta$ with the optical axis. In the mirror reference frame, the wavefront coordinates are $(w_s, h_s) =(-p \cos \theta, p \sin \theta) + (x \sin \theta, x \cos \theta)$. It is straightforward to extend the integral propagator (Eq.~\ref{eq:discreteRSpropagator}) to calculate the propagated field at the surface points $(w,h)$ by summing over all points of the input wavefront. Once the electric perturbation at the mirror surface is known, another propagation is performed using the same principle, to the image plane placed at a distance $q$ from the mirror center.

\section{Examples}

\subsection{A simple example of diffraction limited system}
\begin{figure}[H]
    \centering
    \includegraphics[width=0.99\textwidth]{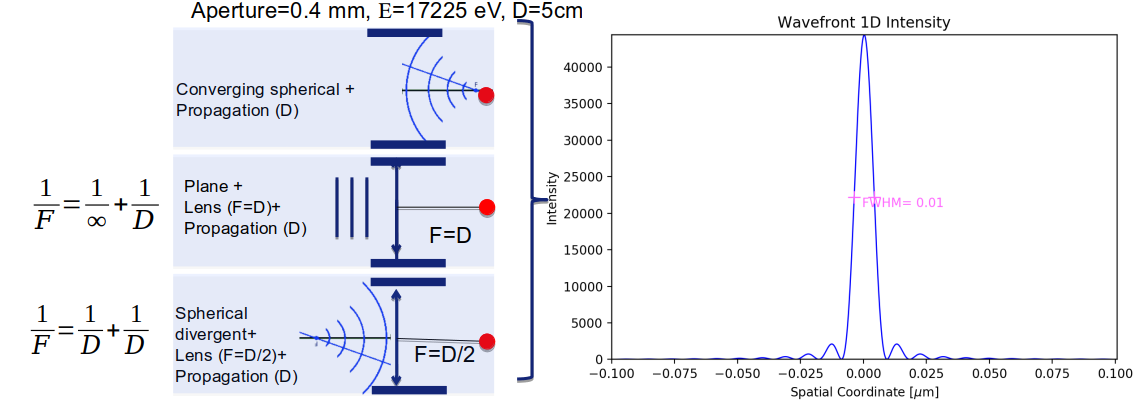}
    \caption{Example of diffraction limited spot of a wavefront focused to a point. Three equivalent systems give the same result: i) a collapsing convergent spherical wavefront (radius D), ii) a plane wave focused by an ideal lens of focal $D$, and iii) a divergent spherical wave (radius R) focused by an ideal lens of focal $D/2$.   }\label{fig:diffractionlimited}
\end{figure}

This short example demonstrates the capabilities of WOFRY to make quick estimation of a diffraction-limited focus. We see in Fig.~\ref{fig:diffractionlimited} a comparison of the three systems that produce identical intensity distributions at a focal point placed at a distance $D$ downstream the element: i) a collapsing spherical wave of radius $D$ limited by a slit, ii) a plane wave focused by an ideal lens of focal $F=D$, and iii) a divergent wave (radius $D$) focused by an ideal lens of focal $F/2$.

\subsection{Changes in the coherent fraction by a ``coherent slit"}
\begin{figure}[H]
    \centering
    \includegraphics[width=0.95\textwidth]{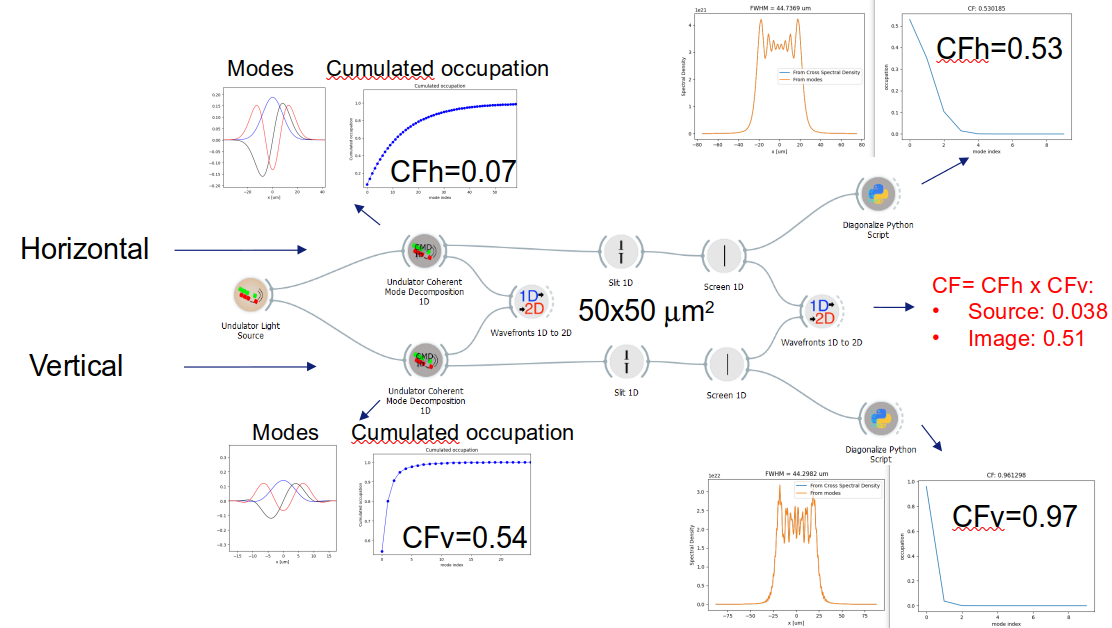}
    \caption{Example of a calculation of coherent fractions with WOFRY for the horizontal and vertical directions (see text).}\label{fig:croppingslit}
\end{figure}

\begin{figure}
    \centering
    \includegraphics[width=0.75\textwidth]{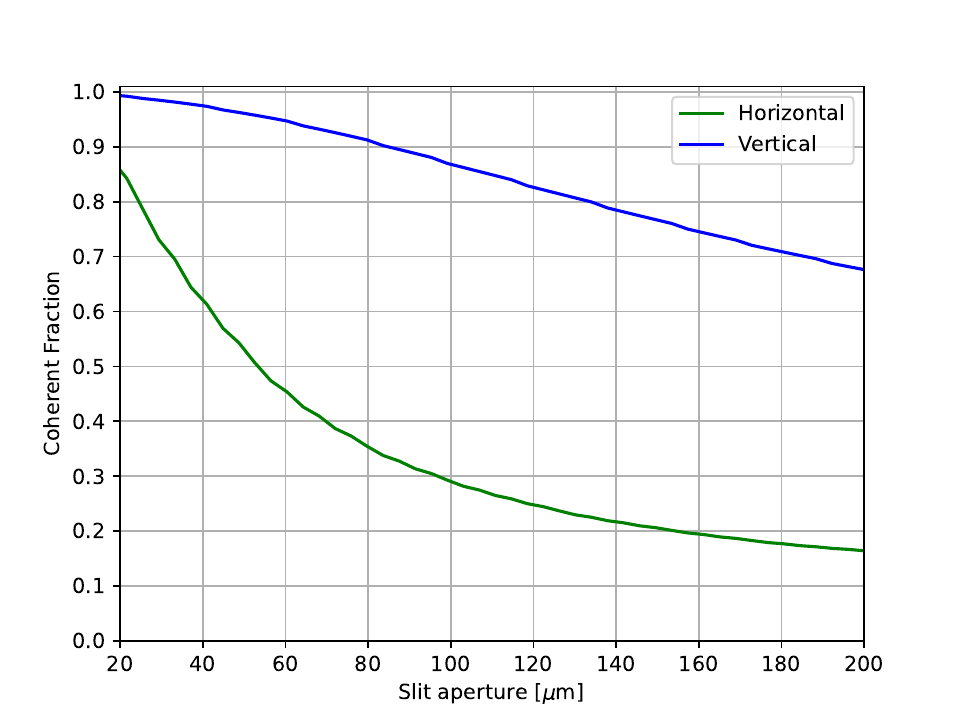}
    \caption{Variation of the coherent fraction versus slit aperture for the system presented in Fig.~\ref{fig:croppingslit}.   }\label{fig:slitscan}
\end{figure}

We want to calculate how the coherent fraction at the source and after a slit. 
We calculated the CMD of a U17 undulator in the EBS storage ring tuned at resonance of 15 keV, resulting a CF of  0.07 in horizontal (H) and 0.54 in vertical (V).  A calculation of the wavefront cropped by a slit of 50~$\mu$m placed a 20~ m is shown in (Fig.~\ref{fig:croppingslit}). The resulting transported modes are used to calculate the propagated CSD and then performing a new CMD (using the WOFRY widget ``Diagonalize Python Script"). After this operation, the resulting CF is 0.53 in H and 0.97 in V, just verifying that a cropping slit increases the beam coherent (the price to pay is the reduction of the intensity). A parametric calculation of the CF and transmission intensity as a function of the slit aperture is done and the results are displayed in Fig.~\ref{fig:slitscan}. This calculation permits to the beamline users to select the desired coherence fraction by selecting the appropriated slit aperture. 

\subsection{Comparison of undulator emission by numeric CMD with approximated GSM}\label{sec:CMDvsGSM}
We compare here the coherent mode decomposition for the U20 undulator (N=100, K=1.19, first harmonic at 10 keV) at the EBS storage ring\footnote{We used the electron beam sizes and divergences 
$\sigma_x$=\SI{30.18}{\micro\meter},
$\sigma_{x'}$=\SI{4.37}{\micro\radian},
$\sigma_y$=\SI{3.64}{\micro\meter},
$\sigma_{y'}$=\SI{1.37}{\micro\radian},
corresponding to beam emittances:  $\varepsilon_x$=\SI{132}{\pico\meter \radian},
$\varepsilon_y$=\SI{5}{\pico\meter \radian}.
}.
We performed CMD of the undulator radiation using the ``Undulator Coherent Mode Decomposition 1D" in Sec.~\ref{sec:CMD} and we also computed the undulator emission using the GSM with the ``Undulator GSM 1D" widget described in Sec.~\ref{sec:undulatorGSM}. 

\begin{figure}
    \centering
    \includegraphics[width=0.95\textwidth]{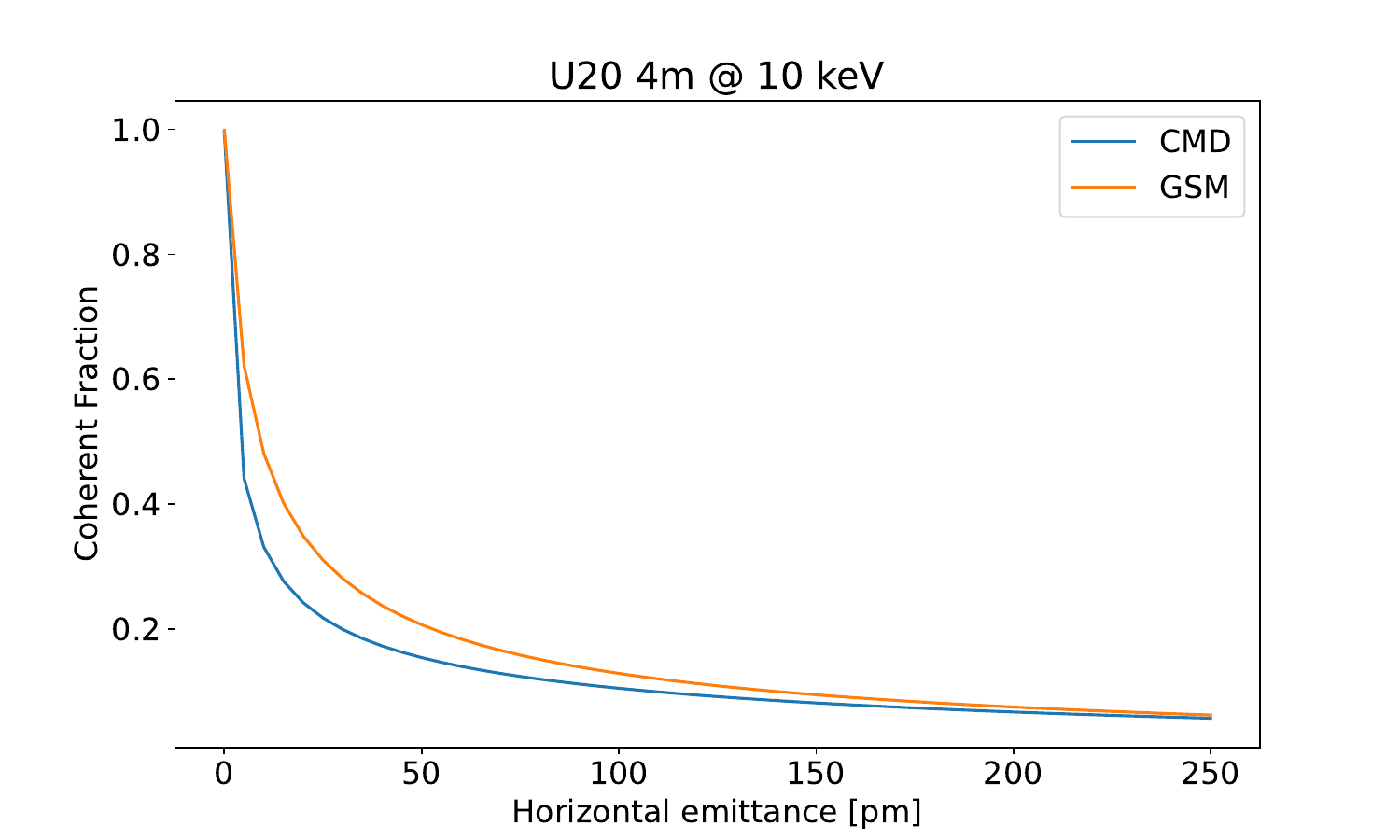}
        
    \caption{Comparison of the coherent fraction calculated using the CMD and approximated value (Eq.~\ref{eq:CFphasespace}) versus emittance.}
    \label{fig:CFvsEmittance}
\end{figure}

The used method for calculating the GSM parameters matches the approximated undulator coherent fraction (Eq.~\ref{eq:CFphasespace}). The approximated CF approaches well the more accurate CMD for ``intermediate" emittance values. Fig.~\ref{fig:CFvsEmittance} compares the CF values given by the CMD with those given by Eq.~\ref{eq:CFphasespace}. It  is appreciated significative differences in the interval 5-50 pm. Indeed, for the values in use (132 pm in horizontal and 5 pm in vertical), the CF in horizontal are 0.09 (CMD) and 0.10 (GSM), showing a good agreement, but they discrepate in about 20\% in vertical: 0.60 (CMD) and 0.75 (GSM). 

\begin{figure}[H]
    \centering
~~~~~~~~~~~~~~~~~~~a1)~~~~~~~~~~~~~~~~~~~~~~~~~~~~~~~~~~~~~~~~~a2)~~~~~~~~~~~~~~~~~~~~~~~~~~~~~~~~~~~~~~~~~~~~~a3)~~~~~~~~~~~~~~~~~~~~~~~~~~~~~~~~~~~~~~~~~\\
    \includegraphics[width=0.25\textwidth]{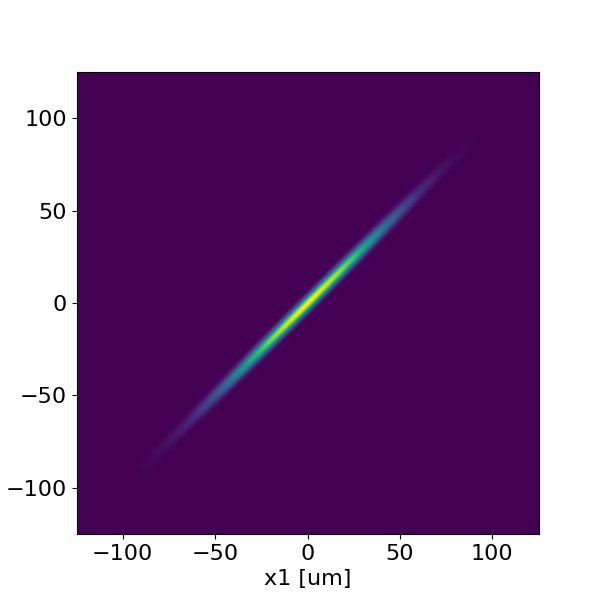}
    \includegraphics[width=0.25\textwidth]{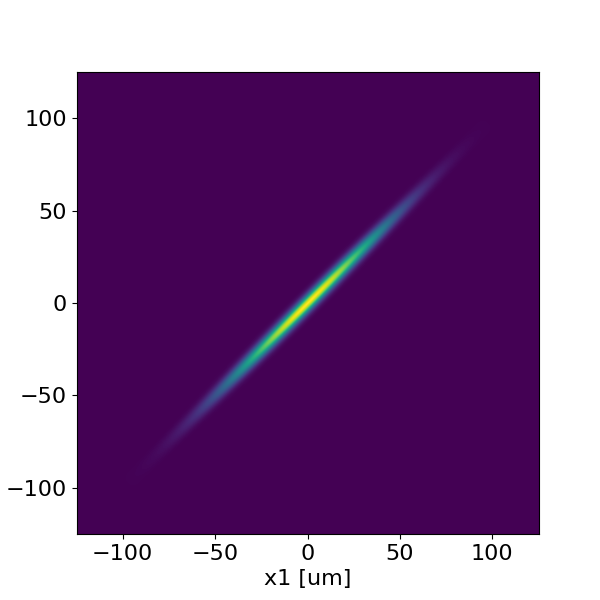}
    \includegraphics[width=0.45\textwidth]{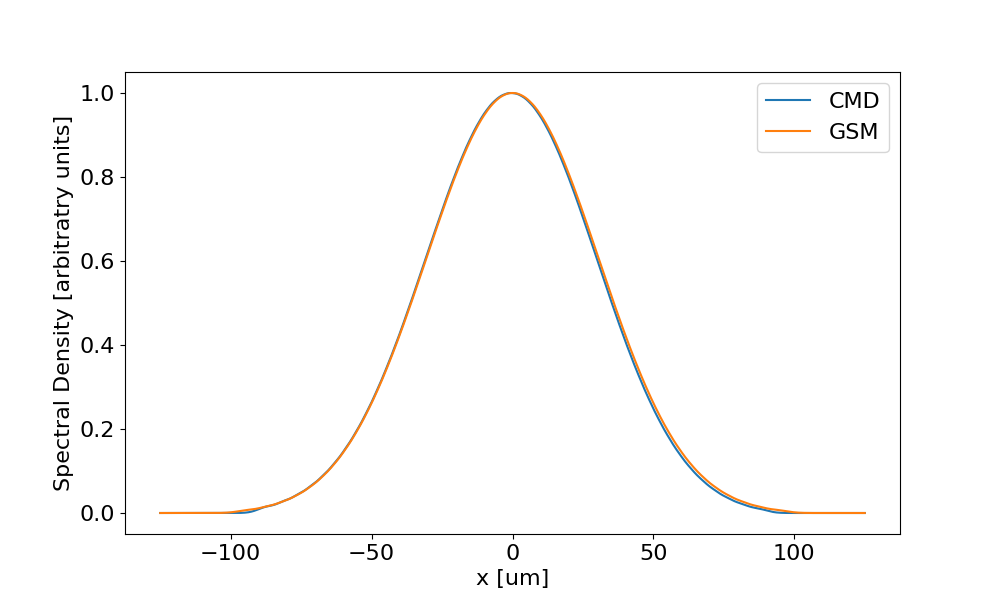}\\
~~~~~~~~~~~~~~~~~~~b1)~~~~~~~~~~~~~~~~~~~~~~~~~~~~~~~~~~~~~~~~~b2)~~~~~~~~~~~~~~~~~~~~~~~~~~~~~~~~~~~~~~~~~~~~~b3)~~~~~~~~~~~~~~~~~~~~~~~~~~~~~~~~~~~~~~~~~\\
    \includegraphics[width=0.25\textwidth]{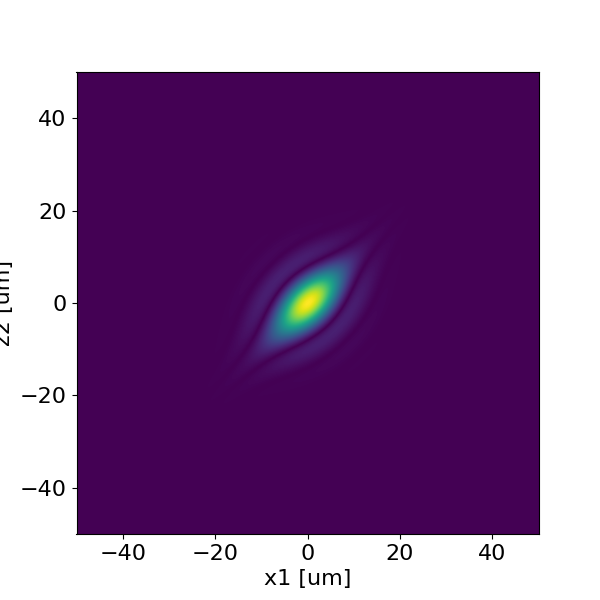}
    \includegraphics[width=0.25\textwidth]{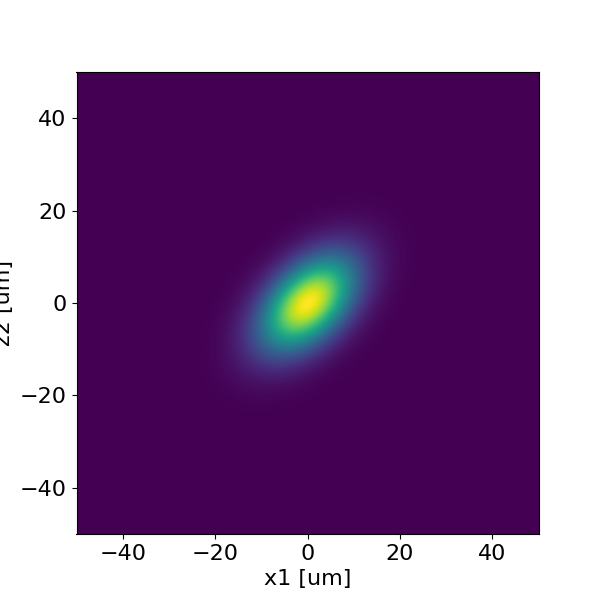}
    \includegraphics[width=0.45\textwidth]{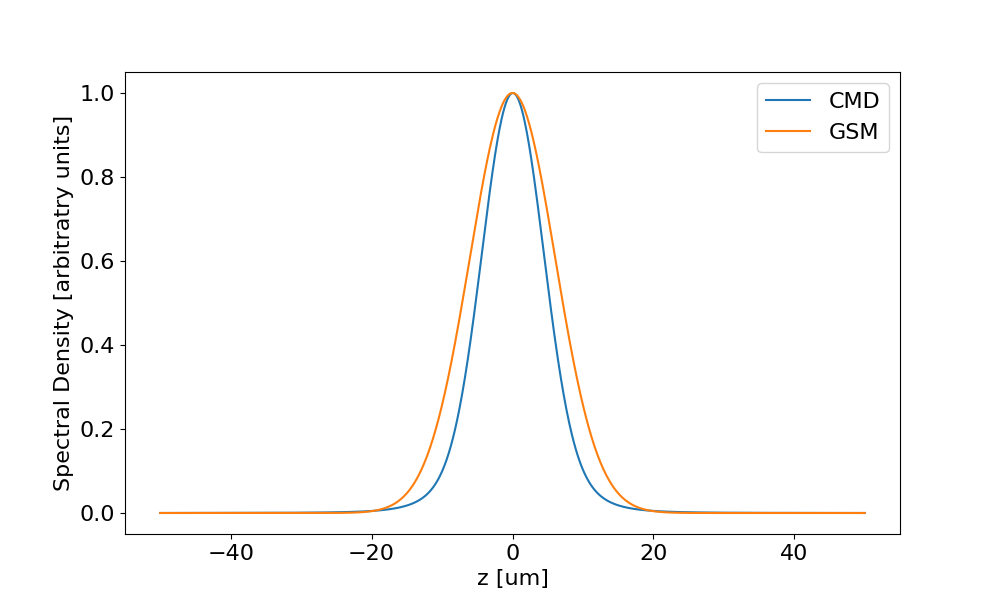}

    \caption{Comparison of the Cross Spectral Density (CSD) and Spectral Density (SD) calculated by numeric coherent mode decomposition and approximated by the Gaussian Shell-model for the horizontal (H) and vertical (V) directions of a U20 undulator of length \SI{2}{\meter} at 10 keV.
    a1) CSD by CDM in H,
    a2) CSD by GSM in H,
    a3) SD in H;
    b1) CSD by CMD in V,
    b2) CSD by GSM in V,
    b3) SD in V.
    }
    \label{fig:GSMvsUND}
\end{figure}

\begin{figure}[H]
    \centering
    ~~~~~~~~~~~~~~~~~~~a1)~~~~~~~~~~~~~~~~~~~~~~~~~~~~~~~~~~~~~~~~~a2)~~~~~~~~~~~~~~~~~~~~~~~~~~~~~~~~~~~~~~~~~~~~~a3)~~~~~~~~~~~~~~~~~~~~~~~~~~~~~~~~~~~~~~~~~\\
    \includegraphics[width=0.25\textwidth]{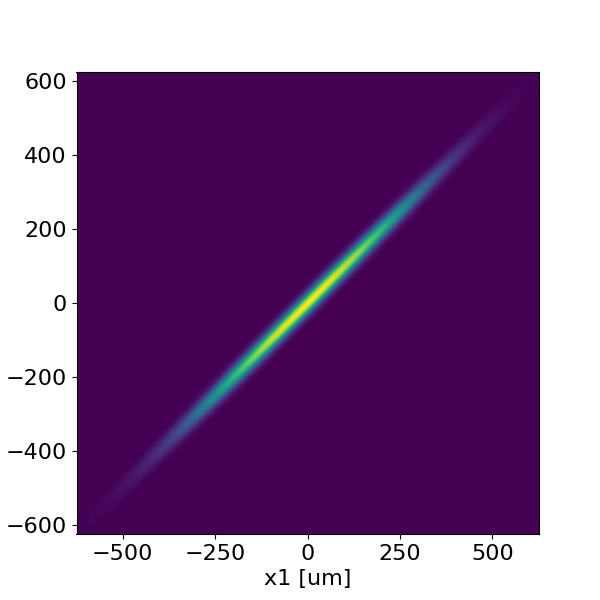}
    \includegraphics[width=0.25\textwidth]{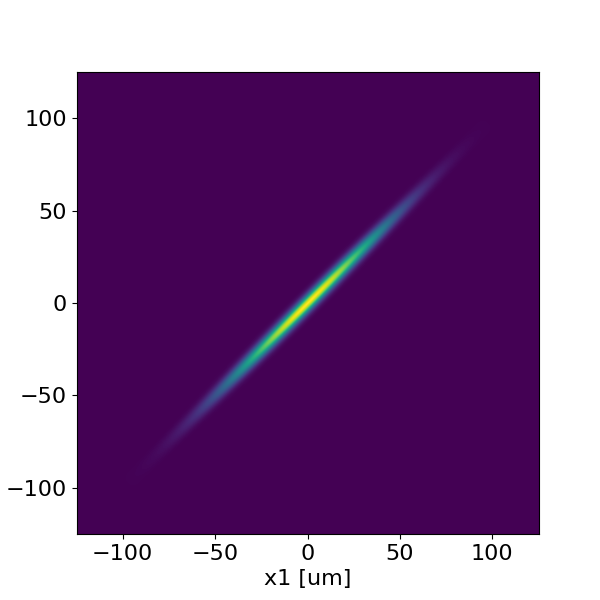}
    \includegraphics[width=0.45\textwidth]{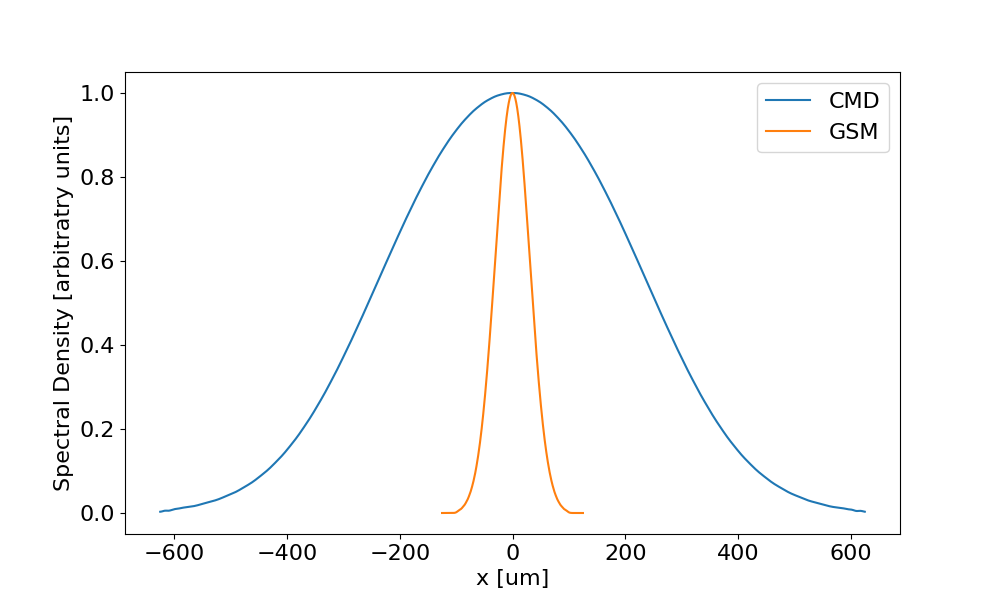}\\
~~~~~~~~~~~~~~~~~~~b1)~~~~~~~~~~~~~~~~~~~~~~~~~~~~~~~~~~~~~~~~~b2)~~~~~~~~~~~~~~~~~~~~~~~~~~~~~~~~~~~~~~~~~~~~~b3)~~~~~~~~~~~~~~~~~~~~~~~~~~~~~~~~~~~~~~~~~\\
    \includegraphics[width=0.25\textwidth]{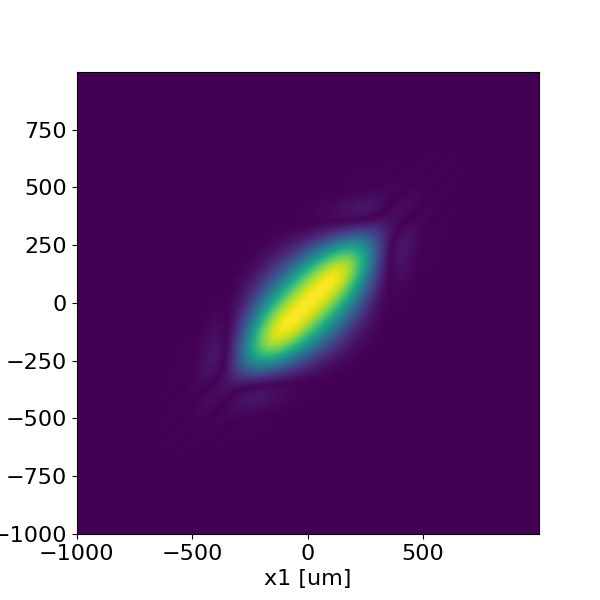}
    \includegraphics[width=0.25\textwidth]{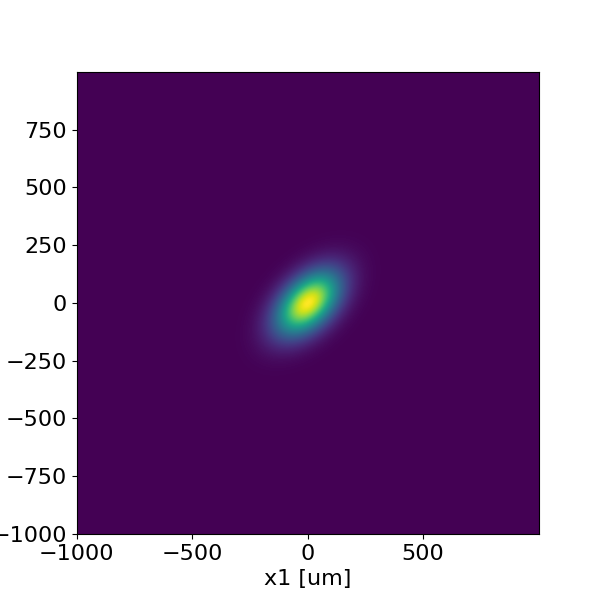}
    \includegraphics[width=0.45\textwidth]{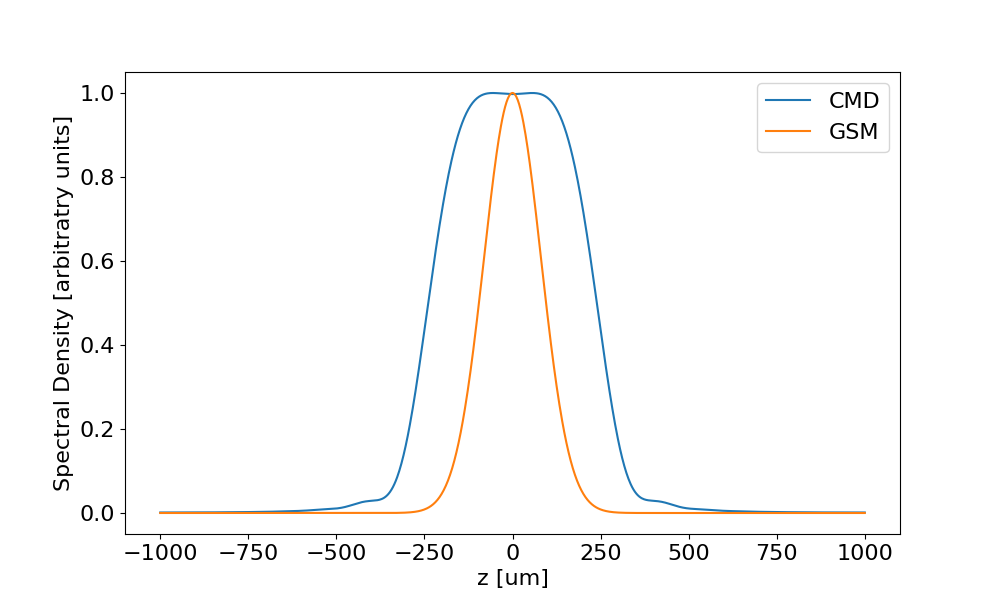}

    \caption{The same as in Fig.~\ref{fig:GSMvsUND} but propagated to a plane placed at \SI{30}{m} downstream from the source.
    }
    \label{fig:GSMvsUND-propagated}
\end{figure}

The spectral densities are similar for the exact and GSM calculation (Fig.~\ref{fig:GSMvsUND}), but it can be appreciated differences in the cross-spectral density for the vertical direction.
When the beams are propagated at a distance \SI{30}{meter} downstream from the source, a large disagreement between the exact and GSM is found (Fig.~\ref{fig:GSMvsUND-propagated}). This is certainly due to the fact that the propagation of the undulator beam is very different than the propagation of the GSM, even if, as shown in Fig.~\ref{sec:CMDvsGSM}, the beams at the source present a reasonably good agreement. The GSM would work only for cases where the coherence fraction is very small like in the old generations of synchrotron sources.

\subsection{Other systems calculated with WOFRY}

WOFRY has played a crucial role in numerous studies, also contributing to the enhancement and optimization of the code and algorithms. Examples include analyzing the efficacy of adaptive X-ray optics in correcting mirror deformations induced by heat load \citep{SanchezdelRio2020}, studying modifications in beam focus due to cropping partially coherent X-ray beams \citep{SanchezdelRio2022EPL}, and assessing beam quality after reflection by a toroidal mirror in a 4th generation storage-ring beamline \citep{ReyesHerrera2023}.

\section{Summary and conclusions}
\label{sec:summary}
In this paper we describe the models implemented in WOFRY (Wave Optics FRamework in pYthon), a wave optics toolbox. It implements a useful model for computing the coherence parameters of synchrotron beamlines, like cross-spectral density and coherent fraction. The software is included in the OASYS \citep{OASYS} suite. We have illustrated the applicability of WOFRY with some examples, from simple diffraction limited waves, to computing the coherent fraction using ``coherent slits". Moreover WOFRY also implements the undulator radiation approximated by the Gaussian Shell-model. In section~\ref{ec:CMDvsGSM} we compare this method with the more exact undulator radiation by numeric coherent mode decomposition. It is shown that, although GSM may reproduce reasonably well the beam characteristics at the source position, but after propagation the differences become important. This

\bibliographystyle{agsm}

\bibliography{iucr} 

@article{OASYS,
  doi = {10.1117/12.2274263},
  url = {https://doi.org/10.1117/12.2274263},
  year = {2017},
  month = aug,
  publisher = {{SPIE}},
  author = {Luca Rebuffi and Manuel {Sanchez del Rio}},
  editor = {Kawal Sawhney and Oleg Chubar},
  title = {{OASYS} ({OrAnge} {SYnchrotron} Suite): an open-source graphical environment for x-ray virtual experiments},
  journal = {{SPIE} Proceedings},
  volume = {10388, \textit{Advances in Computational Methods for X-Ray Optics {IV}}},
}

@inproceedings{syned,
  doi = {10.1117/12.2274232},
  url = {https://doi.org/10.1117/12.2274232},
  year = {2017},
  month = aug,
  publisher = {{SPIE}},
  author = {Luca Rebuffi and Manuel {Sanchez del Rio}},
  editor = {Kawal Sawhney and Oleg Chubar},
  title = {Interoperability and complementarity of simulation tools for beamline design in the {OASYS} environment},
  booktitle = {Advances in Computational Methods for X-Ray Optics {IV}}
}

@inproceedings{XRT,
  title = {Powerful scriptable ray tracing package xrt},
  ISSN = {0277-786X},
  url = {http://dx.doi.org/10.1117/12.2061400},
  DOI = {10.1117/12.2061400},
  booktitle = {Advances in Computational Methods for X-Ray Optics III},
  publisher = {SPIE},
  author = {Klementiev,  Konstantin and Chernikov,  Roman},
  editor = {Sanchez del Rio,  Manuel and Chubar,  Oleg},
  year = {2014},
  month = sep 
}

@article{SanchezdelRio2020,
author = "{Sanchez del Rio}, Manuel and Wojdyla, Antoine and Goldberg, Kenneth A. and Cutler, Grant D. and Cocco, Daniele and Padmore, Howard A.",
title = "{Compensation of heat load deformations using adaptive optics for the ALS upgrade: a wave optics study}",
journal = "Journal of Synchrotron Radiation",
year = "2020",
volume = "27",
number = "5",
pages = "1141--1152",
month = "Sep",
doi = {10.1107/S1600577520009522},
url = {https://doi.org/10.1107/S1600577520009522},
}

@article{SanchezdelRio2022EPL,
doi = {10.1209/0295-5075/aca4ef},
url = {https://dx.doi.org/10.1209/0295-5075/aca4ef},
year = {2022},
month = {dec},
publisher = {EDP Sciences, IOP Publishing and Società Italiana di Fisica},
volume = {140},
number = {5},
pages = {55001},
author = {Manuel {Sanchez del Rio} and Rafael Celestre and Juan Reyes-Herrera and Philipp Brumund and Marco Cammarata},
title = {Beam focus modifications by cropping partially coherent X-ray beams},
journal = {Europhysics Letters},
}

@article{SanchezdelRio2022CMD,
author = "Sanchez del Rio, Manuel and Celestre, Rafael and Reyes-Herrera, Juan and Brumund, Philipp and Cammarata, Marco",
title = "{A fast and lightweight tool for partially coherent beamline simulations in fourth-generation storage rings based on coherent mode decomposition}",
journal = "Journal of Synchrotron Radiation",
year = "2022",
volume = "29",
number = "6",
pages = "1354--1367",
month = "Nov",
doi = {10.1107/S1600577522008736},
url = {https://doi.org/10.1107/S1600577522008736},
}

@Article{ReyesHerrera2023,
AUTHOR = { Reyes-Herrera, J and Celestre, R and Cammarata, M and Barrett, R and Levantino, M and Sanchez del Rio, M},
TITLE = {X-ray beam quality after a mirror reflection: Experimental and simulated results for a toroidal mirror in a 4th generation storage-ring beamline [version 1; peer review: 2 approved]
},
JOURNAL = {Open Research Europe},
VOLUME = {3},
YEAR = {2023},
NUMBER = {173},
DOI = {10.12688/openreseurope.16211.1}
}

@article{Celestre2019,
author = "Celestre, Rafael and Berujon, Sebastien and Roth, Thomas and Sanchez del Rio, Manuel and Barrett, Raymond",
title = "{Modelling phase imperfections in compound refractive lenses}",
journal = "Journal of Synchrotron Radiation",
year = "2020",
volume = "27",
number = "2",
pages = "305--318",
month = "Mar",
doi = {10.1107/S1600577519017235},
url = {https://doi.org/10.1107/S1600577519017235},
}

@book{elleaume,
  title={Undulators, Wigglers and Their Applications},
  author={Onuki, H. and Elleaume, P.},
  isbn={9780203218235},
  year={2003},
  publisher={CRC Press}
}

@book{mandel_wolf,
  title={Optical Coherence and Quantum Optics},
  author={Mandel, L. and Wolf, E.},
  isbn={9780521417112},
  lccn={93488873},
  year={1995},
  publisher={Cambridge University Press}
}

@article{coisson1997,
author = "Co{\"\i}sson, R. and Marchesini, S.",
title = "{Gauss{--}Schell Sources as Models for Synchrotron Radiation}",
journal = "J. Synchrotron Radiat.",
year = "1997",
volume = "4",
number = "5",
pages = "263--266",
month = "Sep",
doi = {10.1107/S0909049597008169},
url = {http://doi.org/10.1107/S0909049597008169}
}

@article{Starikov82, 
author = {A. Starikov and E. Wolf}, 
journal = {J. Opt. Soc. Am.}, 
number = {7}, 
pages = {923--928}, 
publisher = {OSA},
title = {Coherent-mode representation of Gaussian Schell-model sources and of their radiation fields}, 
volume = {72}, 
month = {Jul},
year = {1982},
doi = {10.1364/JOSA.72.000923},
url = {http://doi.org/10.1364/JOSA.72.000923},
}

@article{glass2017,
  author={Glass, Mark and Sanchez del Rio, Manuel},
  title={Coherent modes of X-ray beams emitted by undulators in new storage rings},
  journal={EPL (Europhysics Letters)},
  volume={119},
  number={3},
  pages={34004},
  url={https://doi.org/10.1209/0295-5075/119/34004},
  doi = "{10.1209/0295-5075/119/34004}",
  year={2017},
}

@article{codeSRW,
    author = {O. Chubar and P. Elleaume},
    title = {Accurate and efficient computation of synchrotron radiation in the near field region},
    journal = {Proceedings of the 6th European Particle Accelerator Conference - EPAC-98},
    pages = {1177--1179},
    url = {http://accelconf.web.cern.ch/AccelConf/e98/PAPERS/THP01G.PDF},
    year = {1998},
}

@Article{wise,
author = {D. Spiga and D.  Cocco and C. L.  Hardin and D. S. Morton and M. L.  Ng},
title = {Simulating the optical performances of the LCLS bendable mirrors using a 2D physical optics approach},
journal = {Proceedings SPIE},
volume = {10761},
pages = {1076107},
year = {2018},
doi = {10.1117/12.2323253},
URL = {https://doi.org/10.1117/12.2323253},
eprint = {}
}

@mastersthesis{pirro,
  author       = {Giovanni Pirro}, 
  title        = {Applications of Scaled Wave Optics Propagator to Model Synchrotron Beamlines},
  school       = {Politecnico di Milano},
  year         = 2017,
  howpublished = "\url{https://github.com/oasys-kit/wofry/blob/master/doc/pirro_thesis.pdf}"
}

@book{schmidt,
	author = {Schmidt, Jason D.},
	title = {Numerical Simulation of Optical Wave Propagation},
	publisher = {SPIE Press},
	address = {Bellingham, WA, USA}, 
	year = {2010},
}

@Book{goodmanfourier,
  Title                    = {Introduction to Fourier Optics},
  Author                   = {J. W. Goodman},
  Publisher                = {w.h.freeman},
  Year                     = {2017},
  Address                  = {New Yor, USA},
  Edition                  = {Four},
  ISBN                     = {10: 1-319-11916-6}
}

@article{MOI,
  title = {Mutual optical intensity propagation through non-ideal mirrors},
  volume = {24},
  ISSN = {1600-5775},
  url = {http://dx.doi.org/10.1107/S1600577517010281},
  DOI = {10.1107/s1600577517010281},
  number = {5},
  journal = {Journal of Synchrotron Radiation},
  publisher = {International Union of Crystallography (IUCr)},
  author = {Meng,  Xiangyu and Shi,  Xianbo and Wang,  Yong and Reininger,  Ruben and Assoufid,  Lahsen and Tai,  Renzhong},
  year = {2017},
  month = aug,
  pages = {954–962}
}

\end{document}